\documentstyle[12pt]{article}
\newcommand{\extraspace}{\addtolength{\abovedisplayskip}{2mm}
                        \addtolength{\belowdisplayskip}{2mm}
                        \addtolength{\abovedisplayshortskip}{2mm}
                        \addtolength{\belowdisplayshortskip}{2mm}}
\newcommand{\be}{\begin{equation}\extraspace}

\newcommand{\ee}{\end{equation}}

\newcommand{\bea}{\begin{eqnarray}\extraspace}
\newcommand{\beastar}{\begin{eqnarray*}\extraspace}
\newcommand{\eea}{\end{eqnarray}}
\newcommand{\eeastar}{\end{eqnarray*}}

\begin{document}
\baselineskip=17pt

\hfill {ITFA-96-54}


\vskip 1.5cm
\begin{center}

{\Large The Quantum Hall Effect}

\vskip 4mm

{\large Unified Scaling Theory and Quasi-particles at the Edge}

\vskip 1.5cm

{\large A.M.M. Pruisken and K. Schoutens}

\vskip .3cm

{\sl Institute for Theoretical Physics \\
     University of Amsterdam \\
     Valckenierstraat 65 \\
     1018 XE  Amsterdam\\ 
     THE NETHERLANDS}

\vskip .7cm

{\bf Abstract}

\end{center}

\baselineskip=15pt

{\small
\noindent 
We address two fundamental issues in the physics of the quantum 
Hall effect: a unified description of {\em scaling behavior}\ of 
conductances in the integral and fractional regimes, and a 
{\em quasi-particle formulation}\ of the chiral Luttinger Liquids
that describe the dynamics of edge excitations in the fractional
regime.}

\newpage

\section{Introduction and summary}

A unified description of the scaling behavior in the integral and 
fractional quantum Hall regimes requires a treatment that combines the 
effects of {\em impurities}\ and of {\em electron-electron}\ 
interactions. There are two recent developments that have cleared 
the way for a serious effort in this direction. The first is 
the use of Chern-Simons gauge theory, which makes it possible 
to construct a mapping between the integral and fractional regimes. 
The second development is the recent analysis by Pruisken and 
Baranov (1995) of scaling behavior in the presence of the Coulomb 
interaction. In this paper we briefly review these results.

Effective theories for excitations at the edge of a quantum Hall
sample take the form of a chiral Fermi liquid (in the integral 
regime) or chiral Luttinger liquid (in the fractional regime). 
The Luttinger liquid behavior of edge currents in quantum Hall 
samples with filling fraction $\nu = 1/3$ has been probed in
recent experiments (Milliken et al 1996,  Chang et al 1996). 
In recent theoretical work, the Conformal 
Field Theories that describe the chiral Luttinger liquids for filling 
fractions $\nu=1/(2m+1)$ have been analyzed using a so-called 
quasi-particle formulation. We shall here briefly explain that 
many of the remarkable features of the fractional quantum Hall 
edge dynamics are naturally explained in a such a quasi-particle 
picture.

\section{The quest for a unified scaling theory}

Our microscopic understanding of the quantum Hall effect (qHe) has largely 
developed along two separate pathways which for a long time appeared to have 
very little in common (Prange and Girvin 1990). The first and possibly most 
popularly studied route---initiated by Laughlin---is that of the ``clean'' 
states of the 2D electron gas. 
The focus is primarily on the effects of strong correlation between the 
electrons. The fractional qHe, which manifests itself only in high
quality, high mobility heterostructures (Chang 1990), is generally believed to
be such a strongly correlated phenomenon with novel features such as
fractional statistics and charge of quasi-particles.

 The second approach---which so far has been restricted to the regime
of the integral qHe---is that of the ``impure'' or disorder dominated
states of the 2D electron gas. Here, the main physical objective is to
understand the phenomenon of Anderson localization of free electrons,
which manifests itself macroscopically through ``scaling'' of the
conductances with varying experimental parameters such as magnetic
field and temperature (Pruisken 1988). The so-called ``scaling theory'' of the
integral qHe (Fig.~1) has generated a substantial amount of
experimental (Wei et al 1988, Koch et al 1991, Engel et al 1993, Hwang et al 
1993) and numerical (Huckenstein et al 1990, Huo et al 1993) results on the 
critical aspects of the ``transitions'' between adjacent quantum Hall plateaus.
Criticality plays a fundamental role in the theory of metals and
insulators in general and here it elucidates the surprising mechanism of
``delocalization'' of the 2D electron gas in strong magnetic fields
(Physics Today 1988).

Despite the fact that the extreme theoretical approaches of ``clean''
and ``disordered'' states were originally formulated in a totally
different language, it is natural to expect that the true,
experimental situation must vary continuously between these extremes,
the precise physical outcome being determined by sample specific
parameters such as amount and type of disorder and
temperature. Several authors (Laughlin et al 1985) 
have elaborated on an important
prediction of the renormalization theory of the integral qHe which
says that the lower portion of the $\sigma_{xx}-\sigma_{xy}$ scaling
diagram is in fact a ``forbidden'' region for free electrons. This
forbidden region, indicated by ``?'' in Fig. 1, was subsequently
recognized as the fractional quantum Hall regime.

Hence, there has been a long standing expectation of a ``unified 
scaling theory'' which simultaneously incorporates the effects of strong 
correlation and disorder, i.e. describes both integral and fractional qHe.

\begin{figure}
\vspace{9.5cm}
\caption{\small 
Scaling diagram of the conductances in the integral quantum Hall regime 
(Pruisken 1988). The arrows indicate the scaling toward large
sample sizes (or low $T$) and the flow lines are periodic
in $\sigma_{xy}$ with period $e^2/h$.
}
\end{figure}

\section{Chern-Simons gauge theory and instanton vacuum}

In this Section we discuss two important recent developments which in
principle can be (and have already been) used in order to substantiate
the abovementioned conjecture of a unified scaling theory.  First,
there is the Chern-Simons (CS) gauge theory approach (Wilczek 1982, Zhang et 
al 1989, Lopez and Fradkin 1991, Kivelson et al 1992, Halperin et al 1993)
which has
become instrumental also in other strongly correlated electron systems
such as the chiral spin liquid and anyon superconductivity. The second
development is a recent analysis of
scaling behavior in the presence of electron-electron interactions
(Pruisken and Baranov 1995).

\begin{enumerate}
\item[(i)] Several authors have exploited the fact that by coupling CS
gauge fields to an electronic current density one produces a theory
with probability amplitudes which are identical to those obtained in
the original theory (i.e. without the gauge fields present), provided
the CS coupling constant or ``statistics angle'' $\theta$ is
chosen to be a multiple of $2\pi$.

The interesting observation is being made (Lopez and Fradkin 1991) 
that the statistical gauge 
fields---at a mean field level---produce
a picture which in all respects is identical 
to the ``composite fermion'' theory of Jain. Within the semiclassical 
approximation (i.e. mean field plus gaussian fluctuations), however, all the 
physics of the Laughlin incompressible liquid state is reproduced. In 
this way, CS gauge theory formally provides a "mapping" between
the integral and fractional quantum Hall plateaus.

Although the semiclassical theory of the CS gauge fields is strictly 
valid only if one assumes a gap in the energy spectrum, the same
procedure  has nevertheless been applied to any possible compressible 
(metallic) state in the problem (Kivelson et al 1992). 
If such an approach can indeed be justified in general, 
it would imply that the scaling behavior of the integral regime can 
formally be mapped onto that of the fractional regime.
An explicit $sl(2,{\bf Z})$ dual symmetry has been obtained 
between the integral quantum Hall regime, where the results of 
the scaling theory apply, and the free-electron-forbidden fractional quantum 
Hall regime, denoted by ``?'' in Fig. 1.

Unfortunately, there exists very little knowledge which would justify the 
validity of semiclassical approximations for compressible or metallic states. 
For instance, although the half-integral quantum Hall effect formally 
appears within the framework of the semiclassical theory, it is known 
(Halperin et al 1993) that the gauge field fluctuations give rise to divergent 
contributions to the quasi-particle propagator, indicating that it 
is difficult to 
extract (either explicit or implicit) knowledge about the infrared 
behavior of the theory. Another, possibly more serious 
complication arises from the fact that the conductance fluctuations 
are very large at the quantum Hall transitions (Cohen and Pruisken 1993, 
1994). This raises fundamental questions about the validity of linear response 
theory as a whole and of semiclassical analyses of the CS gauge 
theory in particular.  A microscopic theory for electronic disorder is 
obviously what is needed in order to completely understand the problems at 
hand and, ultimately, to take full advantage of the CS gauge theory 
approach. 

Much of the problematics, raised in this Section, has already been encountered 
more explicitly in the extensive experimental studies on scaling in the 
integral quantum Hall regime (Pruisken and Wei 1993). That is, detailed 
comparison between the predictions of the free electron renormalization theory 
and the experimental data taken from low mobility heterostructures has provided 
important insight in the possible role played by the electron-electron 
interactions in the problem.

\item[(ii)] This brings us back to one of the longstanding problems in the
theory of the qHe, which is 
whether and how the topological concept of an {\em instanton vacuum}
---which has such a dramatic impact on the localization of free 
electrons (Pruisken 1984, 1987)---has any relevance for a system of 
{\em interacting electrons}. The issue was addressed in a recent detailed 
analysis by Pruisken and Baranov (1995). 
Finkelstein's generalized sigma model theory 
(Finkelstein 1983,1984,1994) was used as a starting point for 
the analysis and adapted to the problem of strong Landau level quantization. 
It has turned out that instanton effects give rise to essentially the same 
scaling diagram for the conductances as was obtained within the free electron 
theory, with one important exception: the temperature also scales 
non-trivially, and this implies a non-trivial dynamical scaling behavior 
in the problem. 
\end{enumerate}

\section{Toward a unified scaling theory}

Although the Coulomb interaction problem turns out to share many
features with the free electron theory of the qHe (asymptotic freedom, 
instantons etc.), it is true that many of the 
{\em physical objectives} of the original, perturbative Finkelstein approach 
have actually remained without an answer.
For one thing, it is unclear how one should 
go beyond the perturbative metallic regime and extend the
theory to include the insulating or localized phase where the physics is 
presumably dominated by the appearance of the Efros-Shklovskii {\em Coulomb 
gap}.

\begin{figure}
\vspace{10cm}
\caption{\small
Scaling diagram for the fractional quantum
Hall regime with $\sigma_{xy} < {1 \over 2}$, corresponding
to the area denoted by (?) in Fig.~1.
}
\end{figure}

Another, intimately related complication arises in the attempt to 
incorporate the Chern-Simons gauge fields in the effective sigma 
model action. For this purpose, one has to  
deal explicitly with the electrodynamic $U(1)$ gauge invariance 
of the theory. This invariance is retained by the Finkelstein 
approach in a rather indirect fashion only, however, and this aspect 
of the problem has largely gone unnoticed in the literature. 
For instance, renormalization group schemes have been proposed and dealt with
but which do not respect the $U(1)$ gauge
invariance (Belitz and Kirkpatrick 1994). 
This leads to serious complications in studying the dynamics of 
the problem and, in the end, it invalidates 
any attempt to decipher the consequences of the theory for the insulating 
phase where perturbation theory is no longer applicable.

A microscopic derivation of the effective Finkelstein action
has become instrumental in providing an answer to some of the abovementioned
problems (this 
statement and the following ones have been justified in detail elsewhere).
We will proceed by summarizing our main results for a fundamental quantity 
in the theory, i.e. the electronic specific heat $C_V$, which   
reveals much of the low energy physics of the interacting, disordered electron
gas near two spatial dimensions.
The following expression
for the energy $\cal E$ (relative to the zero point energy)
has been extracted from weak coupling, perturbative expansions
\be
{\cal E} = \int^\infty_0 d\epsilon \, \epsilon \, n_{be} (T)  
\rho_{qp} (\epsilon) .
\ee
The meaning of the symbols is as follows. The integral is over the
excitation energies relative to the chemical potential. The $n_{be}=
{1\over{e^{\epsilon/ T} -1}}$ denotes the {\em Bose Einstein} distribution
whereas $\rho_{qp}$ stands for the {\em quasi-particle } density of states. 
The low temperature specific heat is obtained as $C_V = \partial {\cal E} /
\partial T$. 

Notice that the appearance of {\em bosonic} quasi-particles in the 
problem is naturally explained by the known instability of the 
interacting electron 
gas toward the formation electron-hole pairs near the Fermi energy. The 
Coulomb gap is now seen to occur in the quasi-particle density of states
$\rho_{qp}$. For instance, approaching the
metal-insulator transition (which occurs for spatial dimensions larger than 
two) from the metallic side, then the $\rho_{qp} (\epsilon)$ 
varies from being constant in $\epsilon$ to the
algebraic behavior $|\epsilon|^\delta$. If one replaces the quantity 
$\rho_{qp}$ by the 'magnetization' and the $\epsilon$ by 'magnetic field'
then this type of behavior is completely analogous to what happens to the
classical Heisenberg ferromagnet in the (low temperature) Goldstone phase. 
The quasi-particle density of states $\rho_{qp}$ therefore shows all the 
characteristics of a conventional order parameter.

It is important to remark that the formal analogy with the Heisenberg
ferromagnet breaks down in the 'disordered' or insulating phase where a new
energy scale is entering the problem which is determined by the strength of 
the (unscreened) Coulomb potential. The reason is that 'irrelevant' 
operators, which may be neglected as far as the critical transition is 
concerned, now start dominating the behavior of the electron gas at low 
temperatures. Essentially the same conclusion holds for the AC conductivity.
In different words, the {\em dynamical}
scaling in the metallic and insulating phases occurs 
separately and distinctly in practice
and the behavior is not simply related as usual.

Application of the theory to the 2D electron gas puts the quantum Hall 
regime in a interesting physical perspective. Experimental advances are 
obviously necessary in order to test the different electronic mechanisms for
(dynamical) scaling. For example, one of the strongest effects on the {\em 
exponents} of the plateau transition is predicted to occur by placing the 2D 
electron gas parallel to a metallic layer. The image charges now produce an 
effective, finite range interaction between the electrons and this should
reduce the numerical value of the temperature exponent by a factor of two 
(Pruisken and Baranov 1995). 	

To conclude this Section, we briefly indicate how the CS gauge
fields, coupled to the sigma model effective action, provide in principle a 
unifying renormalization group theory of integral and fractional regimes.
Fig. 2 illustrates the scaling results obtained from weak coupling 
(perturbative and non-perturbative) analyses. It is interesting to notice that
Fig. 2 differs from the historical guess (Laughlin et al 1985)
by a slight redefinition of the fixed points on the even denominator lines
1/2, 1/4, etc. The difference is fundamental, however, since the unstable
fixed points at $\sigma_{xy} =1/2, 1/4 $ etc. now describe the half integer 
phase (Halperin et al 1993) which serves as a perturbative, weak coupling
regime in the present context.

Our unifying theory for the integer and fractional quantum Hall regimes
implies that there is the highly non-trivial relation between "edge" and "bulk" 
effects and this is a largely open problem as of yet.
Interestingly enough, one has to deal here with two topological
concepts simultaneously (CS gauge theory and instanton vacuum) which
together should relate the disordered quasi-particle system of the bulk
to the Luttinger liquid behavior of the edge states (see Section 5 below).
 
\section{Quasi-particles at the edge}
  
In the past five years, it has been recognized that much of the unusual 
physics of the (fractional or integer) quantum Hall effect is reflected by 
the properties of the so-called edge states, which are localized at the 
physical boundary of a quantum Hall sample.

While the edge states form a one-dimensional Fermi liquid for integer
filling fractions, they show non-Fermi liquid properties in the
fractional regime.  The theoretical prediction (Wen 1990,1992) is that 
the edge states form a chiral Luttinger Liquid with characteristic 
parameters depending on the filling fraction $\nu$. These parameters
are directly related to the Chern Simons gauge theory for the bulk 
degrees of freedom (see Section 3). The adjective ``chiral'' means that 
at a given  boundary the edge excitations can only travel in one direction. 
Due to this, backscattering is not possible. The chiral Luttinger Liquid 
has been predicted to be stable and universal.

To actually test these theoretical predictions, one needs an experiment 
that directly probes the non-trivial properties of the edge states. 
One possibility for this is to create a
so-called point contact between two fractional quantum Hall edges and
to study the tunneling of the edge excitations through this contact as a
function of temperature and of an applied gate voltage. For 
filling fraction $\nu=1/3$, this experiment has been performed 
(Milliken et al 1996) and the experimental results have been claimed 
to be in agreement with theoretical predictions. In another
experiment (Chang et al 1996) the $I-V$ characteristics for a current 
tunneling from a bulk metal into a $\nu=1/3$ edge have been measured. 
In the non-linear regime, the data can be fitted to a power law 
$I \propto V^\alpha$ with $\alpha=2.7 \pm .06 $ for a specific sample.

The chiral Luttinger liquids for the fractional qHe edge dynamics
are examples of so-called Conformal Field Theories.
For the specific case of edge theories for filling fraction
$\nu=1/(2m+1)$, $m=1,2,\ldots$,  this Conformal Field Theory is 
particularly simple. It is usually analyzed in ``bosonized" form, the
organizing principle being the $U(1)$ Kac-Moody algebra
generated by the charge density operator. We here propose that
it is much more natural to analyze this theory directly in terms
of the basic quasi-particles, which are not electrons but quasi-particles
of charge $e/(2m+1)$ and of scaling dimension $x=1/(4m+2)$. The 
edge electron is a composite of $(2m+1)$ of these quasi-particles
and as such it has scaling dimension $(2m+1)^2  \, x = (2m+1)/2$.

The systematics of the quasi-particle formulation of the
edge theory for $\nu=1/(2m+1)$  have recently been worked out 
(Iso 1995,  Schoutens and van Elburg). They are closely 
analogous to the systematics of the so-called spinon formulations
of the Conformal Field Theory for the Heisenberg and Haldane-Shastry
spin chains (Haldane et al 1992, Bouwknegt et al 1994), which is 
formally identical to the case $m=1/2$ of the qHe edge theories.

To illustrate our point of view, we give one simple but important
example, which is the tunneling density of states $A(\epsilon)$.
This quantity equals the spectral density for one-electron
excitations with excess energy $\epsilon$ over the ground state.
In free electron theories $A(\epsilon)$ does not depend on
$\epsilon$. However, let us now consider creating a one-electron
excitation in a theory with $\nu=1/(2m+1)$. This means creating
$(2m+1)$ quasi-particles with the single constraint that the sum
of their energies equals $\epsilon$. Clearly, an expression
for $A(\epsilon)$ will contain $2m$ free integrations over 
energy variables. It will therefore depend on $\epsilon$ as
$A(\epsilon) \propto (\epsilon-\mu)^{2m}$. Our reasoning
here, which confirms the result of (Wen 1990), nicely
illustrates the power of the quasi-particle formulation.

The power law for $A(\epsilon)$ implies non-linear $I-V$
characteristics for electron tunneling. For $\nu=1/3$ the 
quantitative prediction is $I \propto V^3$, which can be 
compared with the Chang et al experiment. 

The exact theoretical analysis (Fendley et al 1995) of the universal 
conductance curve for the Milliken et al experiment made use of 
another quasi-particle basis of the $\nu=1/3$ edge theory. This
basis has its origin in theoretical work (Goshal and Zamolodchikov
1994) on integrable boundary scattering. The precise relation 
among the various quasi-particle formulations that are relevant
for the tunneling experiments is currently under study.

\vskip 6mm

\noindent {\em Acknowledgement.}\\
This work was supported in part by the
foundation FOM of the Netherlands.

\newpage

\noindent{\Large\bf References}

\begin{itemize}

\item[---]
D. Belitz and T.R. Kirkpatrick, Rev. Mod. Phys. {\bf 166}, 261 (1994).
 
\item[---]
P. Bouwknegt, A.W.W.  Ludwig and K. Schoutens, Phys. Lett. 
{\bf 338B}, 448 (1994).

\item[---] A.M. Chang in ``The Quantum Hall Effect'', Eds. R.E. Prange and 
S.M. Girvin (Springer Verlag, Berlin, 1990).

\item[---] A.M. Chang, L.N. Pfeiffer and K.W. West, 
Phys. Rev. Lett. {\bf 77}, 2538 (1996). 

\item[---] M.H. Cohen and A.M.M. Pruisken, AIP Conf. Series {\bf 286},
 205 AIP (1993); Phys. Rev. {\bf B49}, 4593 (1994).

\item[---] L.W. Engel, D. Shakar, C. Kurdak and D.C. Tsui, Phys. Rev. Lett. 
{\bf 71}, 2638 (1993).

\item[---] P. Fendley, A.W.W. Ludwig and H. Saleur, 
Phys. Rev. Lett. {\bf 74}, 3005 (1995).

\item[---] A.M. Finkelstein, JETP Lett. {\bf 37}, 517 (1983); Sov. Phys. JETP
{\bf 59}, 212 (1984); Physica B {\bf 197}, 636 (1994).

\item[---] S. Goshal and A.B. Zamolodchikov, Int. Jour.
Mod. Phys. {\bf A9}, 3841 (1994).

\item[---] F.D.M. Haldane, Z.N.C. Ha, J.C. Talstra, D. Bernard
and  V. Pasquier, Phys. Rev. Lett. {\bf 69}, 2021 (1992).

\item[---] B.I. Halperin, P.A. Lee and M. Read, Phys. Rev. {\bf 47}, 7312 
(1993).

\item[---] B. Huckestein and B. Kramer, Phys. Rev. Lett. {\bf 64}, 1437 (1990).

\item[---] Y. Huo, R.E. Hetzel and R.M. Bhatt, Phys. Rev. Lett. {\bf 70}, 481 
(1993).

\item[---] S.W. Hwang, H.P. Wei, L.W. Engel, D.C. Tsui and A.M.M. Pruisken, 
Phys. Rev. {\bf B48}, 11416 (1993).

\item[---] S. Iso, Nucl. Phys. {\bf B443} 581 (1995).

\item[---] S. Kivelson, S.C. Zhang and D.H. Lee, Phys. Rev {\bf B46},
2223 (1992).

\item[---] S. Koch, R.J. Hang, K. von Klitzing and K. Ploog, Phys. Rev. Lett. 
{\bf 67}, 883 (1991).

\item[---] R.B. Laughlin, M.L. Cohen, J.M. Kosterlitz, H. Levine,
S.B. Libby and A.M.M. Pruisken, Phys. Rev. {\bf B32}, 1311 (1985).

\item[---] A. Lopez and E. Fradkin, Phys. Rev. {\bf B44}, 5246 (1991).

\item[---] F.P. Milliken, C.P. Umbach and R.A. Webb,  Solid State Comm.
{\bf 97}, 309 (1996).

\item[---] Physics Today, Search \& Discovery, September 1988.

\item[---] ``The Quantum Hall Effect'', Eds. R.E. Prange and S.M. Girvin 
(Springer Verlag, Berlin, 1990). 

\item[---] A.M.M. Pruisken, Nucl. Phys. {\bf B235} [FS 11], 277 (1984).

\item[---] A.M.M. Pruisken, Nucl. Phys. {\bf B285} [FS19], 719 (1987);
Nucl. Phys. {\bf B290} [FS20], 61 (1987).

\item[---] A.M.M. Pruisken, Phys. Rev. Lett. {\bf 61}, 1298 (1988) and 
references therein.

\item[---] a brief account of this work has appeared in A.M.M. Pruisken 
and M.A. Baranov, Europhys. Lett. {\bf 31}, 543 (1995).

\item[---] A.M.M. Pruisken and H.P. Wei, AIP Conf. Series {\bf 286}, 
215 AIP (1993). 

\item[---] K. Schoutens and R. van Elburg, in preparation

\item[---] H.P. Wei, D.C. Tsui, M.A. Palaanen and A.M.M. Pruisken, Phys. Rev. 
Lett. {\bf 61}, 1294 (1988) and references therein.

\item[---] X.G. Wen, Phys. Rev {\bf B41}, 12838 (1990); Int. Jour. Mod. Phys.
{\bf B6}, 1711 (1992).

\item[---] F. Wilczek, Phys. Rev. Lett. {\bf 48}, 1144 (1982).

\item[---] S.C. Zhang, T. Hansson and S. Kivelson, 
Phys. Rev. Lett. {\bf 62}, 82 (1989).

\end{itemize}

\end{document}